\documentclass[aps,prl,twocolumn,superscriptaddress]{revtex4}

\usepackage {graphicx}
\usepackage{amsmath}
\usepackage{epstopdf}

\def\blp{\bigg(}
\def\brp{\bigg)}
\def\la{\langle}
\def\ra{\rangle}

\def\D{\Delta}
\def\d{\delta}
\def\e{\epsilon}

\def\r{\rho}

\begin{document}

\title{Generalized Erd\"os Numbers}
\author{Greg Morrison$^1$ and L. Mahadevan}
\affiliation{School of Engineering and Applied Sciences, Harvard University, Cambridge MA 02138}

\begin{abstract}

We propose a simple real-valued generalization of the well known integer-valued Erd\"os number as a topological, non-metric measure of the `closeness' felt between two nodes in an undirected, weighted 
graph.  These real-valued Erd\"os numbers are asymmetric  and are able to distinguish between network topologies that standard distance metrics view as identical. We use this measure 
to study some simple analytically tractable networks, and show the utility of our measure to devise a ratings scheme based on the generalized Erd\"os number that we deploy on the data from 
the NetFlix prize, and find a significant improvement in our ratings prediction over a baseline.   

\end{abstract}

\maketitle
\date{today}

A variety of complex natural and artificial systems can be viewed as a network \cite{NewmanBook}, with a set of nodes representing objects and a set 
of edges connecting these nodes representing interactions between objects.   Such systems include protein \cite{KingBioinf04} or metabolic \cite{JeongNature00,PatilPNAS05} networks, 
computer networks and the world wide web \cite{ilprints422,AdemicScience00}, disease propagation in populations \cite{KeelingPRSocLond99,SatorrasPRL01}, and networks of human
\cite{TraversSociom69,KeelingPRSocLond99} or other animal \cite{McRaeEvol2006,ProulxTrendsEcolEvol05} interactions. While much of the study of networks generally involves characterizing 
both its internal structure \cite{TraversSociom69,JeongNature00,SatorrasPRL01} and the propagation of dynamical processes in it  \cite{NewmanBook, AlbertNature00}, a 
basic question that continues to be of interest is that of characterizing closeness or connectedness in such networks. Various measures of the distance between nodes have been developed 
including the integer distance \cite{BuckleyBook} (identical to the classic Erd\"os numbers \cite{DeCastroMathIntel99} which measure the authorship-distance to the famous Hungarian 
mathematician) or resistance-distance approaches \cite{KleinJMathChem93,McRaeEvol2006} which often have both a geometric and a topological character to them.  In this paper, we develop a framework for determining the `closeness' between nodes in a weighted network by developing a generalized real-valued Erd\"os number, an inherently topological entity that 
incorporates nonlocal information about connectivity, is asymmetric, i.e. $E_{ij}\ne E_{ji}$ even if the underlying adjacency matrix is symmetric.  Using analytically tractable symmetric networks, 
we show that these Erd\"os numbers can distinguish between topologies that are identical when viewed through the lens of common distance metrics \cite{BuckleyBook,KleinJMathChem93}.  In 
order to show that these Erd\"os numbers have utility in making quantitative predictions about real-world networks, we also develop a basic predictor for a small subset of the NetFlix data
\cite{NetflixPrize1}, and find significant improvement over a baseline prediction.

\begin{figure}[tbp]
\centering
\includegraphics[width=.48\textwidth]{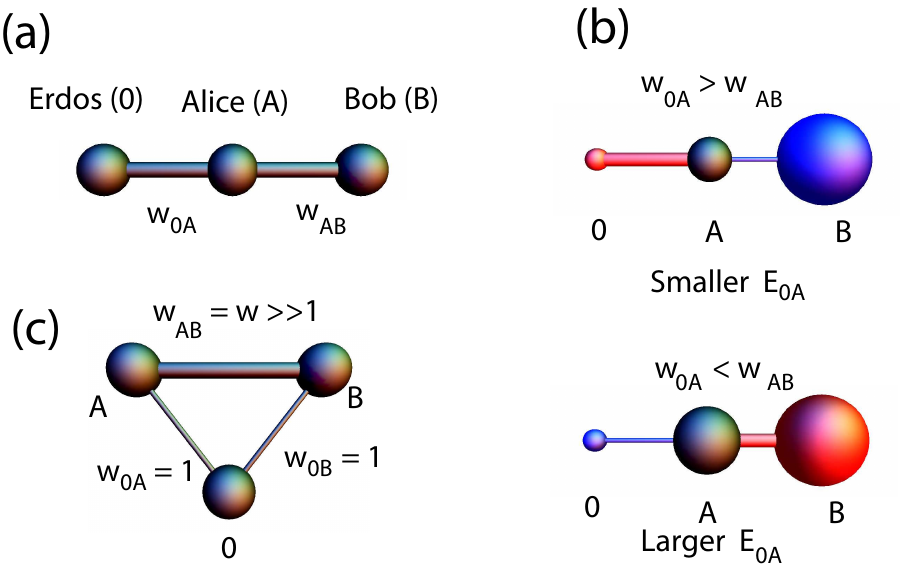}
\caption{(a)  A simple linear author network.  Bob has published only with Alice, so $E_{0B}=E_{0A}+w_{0A}^{-1}$.  The competition between the connections with Erd\"os and leads to $E_{0A}$ 
given in (\ref{LineSolution}).  (b)  Smaller sizes denote closeness to Erd\"os (large spheres implying large $E_{0i}$).  Red coloring denotes strong interactions, while blue denotes weak 
interactions.  $E_{0A}$ increases as $w_{AB}$ increases.  (c)  A simple cycle, for two authors weakly connected to Erd\"os, but strongly connected to one another.  }
\label{SimpleErdosLine.fig}
\end{figure}

In order to develop a natural measure of the `closeness' between two nodes, we consider one of the simplest possible networks:  a linear network of exactly three nodes (diagrammed in Fig. 
\ref{SimpleErdosLine.fig}(a)).  With Erd\"os indexed as 0, we define his closeness to himself as $E_{00}=0$, as is the case in all distance metrics\cite{BuckleyBook,KleinJMathChem93}.  For a 
node B (Bob, say) directly connected to exactly one other node (Alice in this case), we define the closeness felt by Bob towards Erd\"os as 
 $E_{0B}=E_{0A}+w_{AB}^{-1}$,
with $w_{AB}$ the weight of the edge joining Alice and Bob.  The determination of $E_{0A}$ is more ambiguous, since Alice is connected to two nodes.  If we were to use a simple integer- 
\cite{BuckleyBook,DeCastroMathIntel99} or resistance-distance measure \cite{KleinJMathChem93} the distance between Erd\"os and Alice would be $R_{0A}=w_{0A}^{-1}$, whereas one would 
expect a realistic measure of closeness would depend on {\em{all}} of the nodes to which Alice is connected. 

To incorporate the effect of multiple connections between Alice and the other nodes, we assume that the closeness Alice feels to Erd\"os is a function of the closeness felt by all other nodes 
connected to Alice towards Erd\"os.  In particular, we expect $E_{0A}=f(\{E_{0i}+w_{Ai}^{-1}\})$, where $E_{0i}+w_{Ai}^{-1}$ would be the closeness Alice feels to Erd\"os {{in the absence}} of all 
other interactions.  We expect that the unknown functional form of $f$ should (1) penalize large values of $E_{0i}+w_{Ai}^{-1}$ (i.e. that nodes that feel close to Erd\"os contribute more than 
nodes that feel far from Erd\"os when computing $E_{0A}$), and (2) that nodes with high weight have a higher contribution than those of low weights.  These expectations are diagrammed 
schematically in Fig. 1(b), and suggest the use of a weighted harmonic mean of the form
\begin{eqnarray}
\frac{1}{E_{0A}}=\frac{1}{w_{0A}+w_{AB}}\blp\frac{w_{0A}}{E_{00}+w_{0A}^{-1}}+\frac{w_{AB}}{E_{0B}+w_{AB}^{-1}}\brp\label{LineSolution},
\end{eqnarray}
where the necessity of using the scaled weight $w_{Ai}/(w_{A0}+w_{AB})$ will be addressed below.  We note that although (\ref{LineSolution}) is the simplest and most natural functional form 
that satisfies the constraints above, other forms are certainly possible. Furthermore, the centrality of Erd\"os in any network may clearly be replaced by that of any node $i$, so that we
can generalize (\ref{LineSolution}) to define the closeness felt by node $j$ towards node $i$ as
\begin{eqnarray}
\frac{1}{E_{ij}}=\frac{1}{d_j}\sum_{l\in C_j}\frac{w_{jl}}{E_{il}+w_{jl}^{-1}}\label{EnumCalc}
\end{eqnarray}
where $w_{jl}$ is the weight of the edge between $j$ and $l$, $C_j$ is the set of nodes directly connected to node $j$, and $d_j=\sum_{l}w_{jl}$ is the weighted degree of node $j$.  The 
reason for scaled weights $w_{jl}/d_j$ becomes clearer in (\ref{EnumCalc}):  unscaled weights $w_{jl}$ would imply that node $j$ would have a low Erd\"os number $E_{ij}$ (i.e. feel very close 
to node $i$) by having many connections (large $d_j$), even if these connections led to nodes with high Erd\"os numbers $E_{il}$.  We note that if $w_{jl}=\epsilon\d_{jl_0}$ for some $l_0$, then $E_{jl_0}\sim \epsilon^{-1}\to \infty$ as $\epsilon\to 0$, so as a node with vanishing weight for its only connection to the network
will have its Erd\"os numbers diverge, as it becomes `disconnected'.

To illuminate aspects of the generalized Erd\"os numbers we first consider a simple, three-node cycle shown in Fig. \ref{SimpleErdosLine.fig}(c), where two strongly connected nodes with 
weight $w_{AB}=w\gg 1$ are weakly connected to a third node (indexed 0). Solving (\ref{EnumCalc}) for the Erd\"os numbers $E_{ij}$ for this simple network yields $E_{0A}=E_{0B}\sim (1+
\sqrt{5})/2+O(w^{-1})$ for large $w$, showing that the two nodes move away from the third as their connection strengthens (note that $E_{0i}=1$ for $w=1$).  Nodes A and B move towards 
each other as $w$ increases, 
as can be seen by computing $E_{AB}=E_{BA}\sim w^{-1}+O(w^{-2})$.  The third node has a low degree and is closer to the other nodes than they are to it ($E_{A0}=E_{B0}\sim 1+O(w^{-1})$).  Fig. \ref{SimpleErdosLine.fig}(c) thus displays the inherent asymmetry in the Erd\"os numbers 
($E_{A0}\ne E_{0A}$), indicating that $E_{ij}$ is {\em{not}} a distance metric, but rather an inherently topological measure.

\begin{figure}[tbp]
\begin{center}
\includegraphics[width=.48\textwidth]{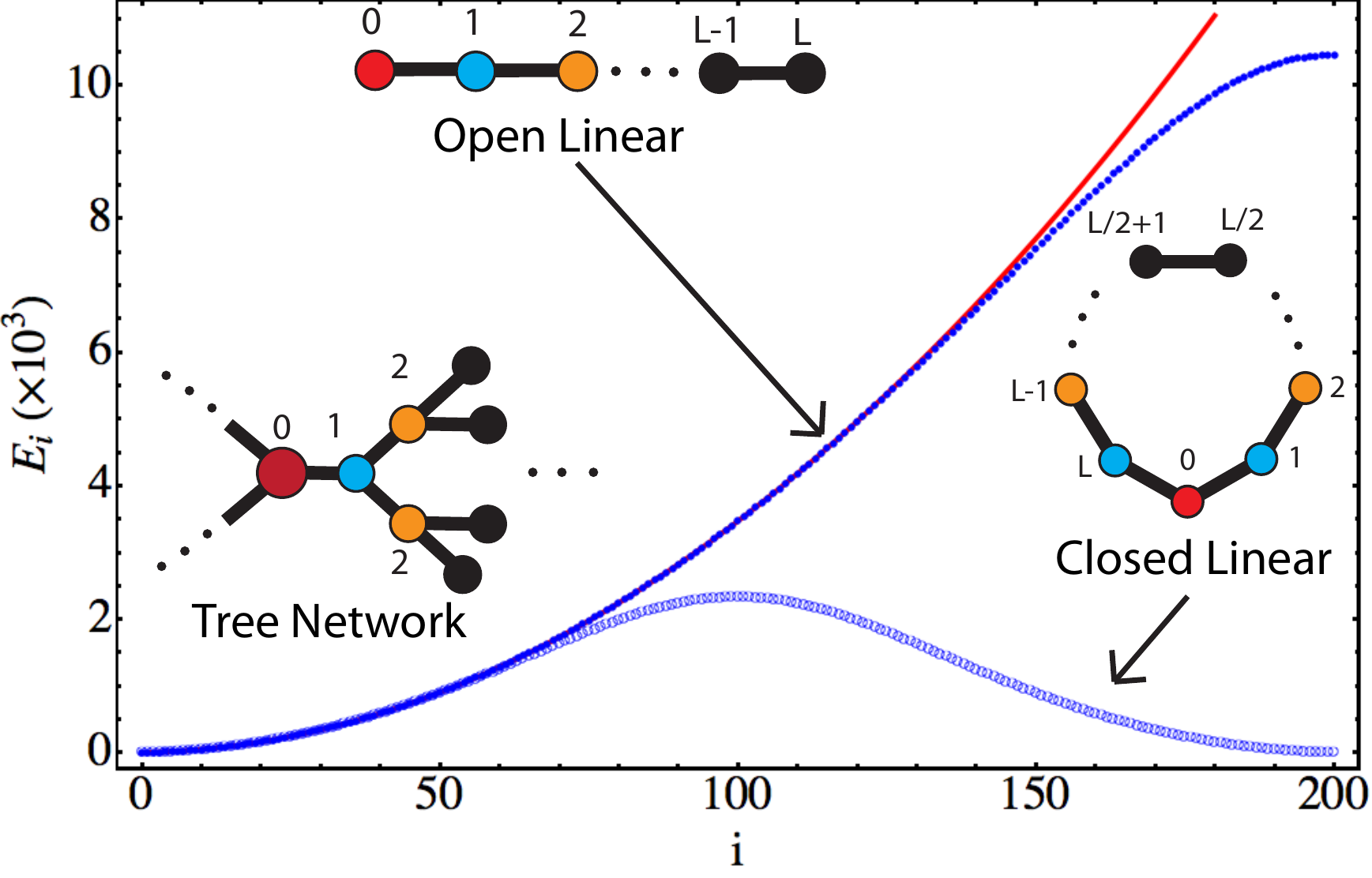}
\caption{
The Erd\"os numbers computed for the open (filled symbols) and closed (open symbols) linear networks, along with the theoretical scaling of $E_{0i}=i(i+4)/3$ (solid red line).  Insets schematically diagram the open and closed linear networks, as well as a tree network with $m=3$ connections per node and length $L=3$ (discussed further in the text).  
}
\label{LargeNetwork.fig}
\end{center}
\end{figure}

The Erd\"os numbers differ from common distance metrics in a number of ways, as can be seen by examining some more complex networks.  In a fully connected network of $N+1$ nodes of constant interaction strength $w_{ij}=w(1-\d_{ij})$, each node has an identical Erd\"os number ($E_{0i}=E$ for $i \ne 0$), with the equivalent of (\ref{EnumCalc}) given by 
\begin{eqnarray}
\frac{wN}{E}=w^2+\frac{w^2(N-1)}{wE+1},
\end{eqnarray}
yielding $E=\sqrt{N}/w$.  As the strength $w$ of each connection between nodes increases, the Erd\"os number of all nodes decrease, since all nodes become closer to each other as well as to
Erd\"os.  However, as the number of nodes increases, with edges added to keep the network fully connected, the importance of an individual edge is lessened and all nodes will feel less close to one another.  This is in contrast with other measures such as the resistance distance, which decreases
as new nodes are added or integer distance, which remains constant independent $N$.  

We next consider generalizations of the simple networks (Fig. \ref{SimpleErdosLine.fig}) to extended linear networks and a cycle-free tree (Fig. \ref{LargeNetwork.fig}) where each node is connected to exactly $m$ nodes, 
except for the endpoints. For the open networks with $m\ge 2$, the resistance distance between any two points is $R_{ij}=|i-j|$, since there are no cycles, while the generalized Erd\"os numbers between a 
node $i$ and the base of a branch are
\begin{eqnarray}
\frac{wm}{E_{0i}}=\frac{w^2}{wE_{0,i-1}+1}+\frac{w^2(m-1)}{wE_{0,i+1}+1}\label{treeEq2}
\end{eqnarray}
with the boundary conditions $E_{00}=0$ and $E_{0L}=E_{0,L-1}+w^{-1}$.  The closed linear network can be studied using the same difference equation, with the boundary conditions 
$E_0=E_{0,L+1}=0$ after insertion of a virtual node.  $E_{0i}(w)=E_{0i}(w=1)/w$ for constant interaction strength $w$, so the weights can be factored out, and are ignored below.  While the difference equations are not exactly solvable,  if $m=2$, we can see that $E_{0i}=i(i+4)/3$ is a solution that satisfies (\ref{treeEq2}) and the boundary condition $E_{00}=0$.  For $i
\approx L$, deviations from this predicted scaling are expected to occur due to the boundary condition at the distant ends. Interestingly, the quadratic scaling $E_{0i}\sim i^2$ 
for distant nodes matches the time for particle diffusion from node 0 to $i$, taking time $\tau\sim i^2$.  For tree networks with large $m$ (inset of Fig. \ref{LargeNetwork.fig}), we find that $E_{0i}=E_{0,i-1}+(m-1)^i$ asymptotically satisfies the difference equation with the boundary condition $E_{00}=0$.  
The tree network produces an exponential growth with $i$ for large $m$, rather than the quadratic growth seen for $m=2$, clearly showing that the Erd\"os numbers are able to distinguish between the global topology of these very different network more accurately than a resistance distance approach.

In order to determine the numerical values of the Erd\"os numbers for this linear network (with $w=1$), we determine an iterative solution for $E_{0i}$, with $2/E_{0i}^{(t)}=(E_{0,i-1}^{(t-1)}+1)^{-1}+(E_{0,i+1}^{(t-1)}+1)^{-1}$ and $E_{0i}^{(0)}=i(i+4)/3$.  $E_{0i}^{(t)}$ is computed until $\e(t)=\max_j |E_{0i}^{(t)}-E_{0,i}^{(t-1)}|<0.01$.  The resulting numerical solutions to the Erd\"os numbers are shown in Fig. \ref{LargeNetwork.fig}, with the solid red line denoting the predicted quadratic growth, $E_{0i}=i(i+4)/3$.  The predicted scaling agrees well with the numerical results \cite{NoteAboutTree}, with deviations occuring near the $i=N$ endpoint for the open network and near the $i=N/2$ midpoint for the closed network.

To see if the Erd\"os numbers can make quantitative predictions about real-world networks, we consider the data provided for the NetFlix Prize \cite{NetflixPrize1}, a competition to improve algorithms for the prediction of movie ratings.  Here, we use the generalized Erd\"os numbers as a means to characterize an interaction `energy' between nodes when predicting the rating user $i$ gives to movie $l$, $p_i^{(l)}$, using the Boltzmann weighted average taken from statistical mechanics,
\begin{eqnarray}
p_i^{(l)}=\sum_{j\in S_l} r_j^{(l)}\ e^{-\beta E_{ij}}\ \bigg/\ \sum_{j\in S_l} e^{-\beta E_{ij}}\label{predictEq}.
\end{eqnarray}
$\beta$ is a free parameter (an inverse temperature), describing how important distant nodes are in determining the predicted rating.  $\beta E_{ij}$ determines which nodes are important to the average and which are not, and assigns a lower weight to the latter.  In order to compute the Erd\"os numbers in (\ref{predictEq}), we 
need to generate a weighted graph use the NetFlix data.

\begin{table}[tdp]
\begin{center}
\begin{tabular}{|c|c|c|c|c|c|}
\hline
case &$N$ & $k$ & $\alpha$ & num. with $n> 30$ & $\kappa_{\beta=2}(n_{min}=30)$\\
\hline
1&3000 & 553 & 2 & 304  (55\%) & 3.56\%\\
\hline
2&3000 & 557 & 4 & 302  (54\%) & 4.71\%\\
\hline
3&3000 & 1297 & 8 & 782  (60\%) & 3.35\%\\
\hline
4&6000 & 368 & 8 & 188 (51\%) & 4.53\%\\
\hline
\end{tabular}
\end{center}
\label{default}
\caption{Parameters used in the NetFlix analysis.  $N$ is the number of users in the dataset, and $k$ is the number of users for whom predictions were made.  The number of nodes with $n_i
\ge n_{min}$ out of the $k$ considered are shown, as well as the average percent improvement for these nodes.}
\end{table}

\begin{figure}[tbp]
\begin{center}
\includegraphics[width=.48\textwidth]{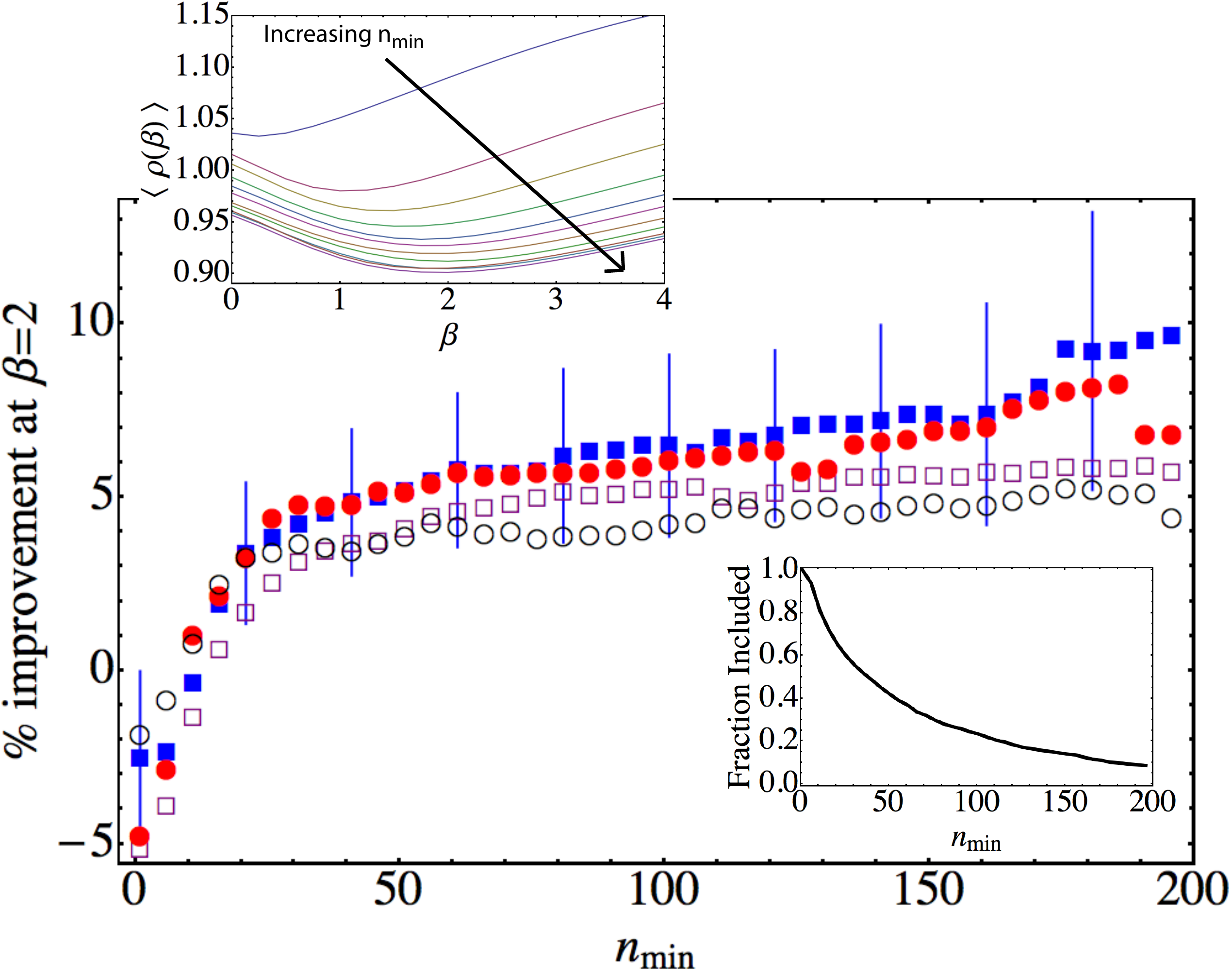}
\end{center}
\caption{
Percent improvement at $\beta=2$ compared to $\beta=0$ as a function of $n_{min}$.  See Table 1 for parameters.  Case 1 is shown as open circles, case 2 as filled circles, case 3 as open squares, and case 4 as filled squares.  Error bars (using the standard deviation of the mean) are shown only for case 4, with the errors for the other cases being smaller.  Upper inset shows $\la \rho(\beta)\ra$ as a function of $\beta$ for varying $n_{min}$ (higher curves correspond to smaller $n_{min}$.  The lower inset shows the fraction of users satisfying $n\ge n_{min}$.
}
\label{RMSD.fig}
\end{figure}

While we could represent the NetFlix data as a bipartite network \cite{ZhouPNAS10}, where the users and movies form sets of disjoint nodes, we instead use the movie ratings (an integer between 1 and 5) to determine a weight between two users, using the simple power law form 
\begin{eqnarray}
w_{ij}=\sum_{l\in M_{ij}} (5-|\Delta r_{ij}^{(l)}|)^\alpha\label{weightEq}
\end{eqnarray}
with $\D r_{ij}^{(l)}=r_i^{(l)}-r_j^{(l)}$, $r_i^{(l)}$ the rating user $i$ gave to movie $l$ ($0\le |\D r_{ij}^{(l)}|\le 4$), and $M_{ij}$ is the set of movies that both user $i$ 
and $j$ have rated ($w_{ij}=0$ if $i$ and $j$ have rated no movies in common).  If users $i$ and $j$ disagree on all movies (i.e. one rates a 5 while the other rates a 1), the weight between them is $w_{ij}=|M_{ij}|$, while perfect agreement gives a weight $w_{ij}=5^\alpha
\times|M_{ij}|$.  Implicit in this definition is that users who seek out the same movies have more similar tastes than those who do not (even if they do not agree), and that users who 
agree on movies are more likely to have similar tastes than those who disagree.  The free parameter $\alpha$ determines the importance of agreement, with $\alpha=0$ implying that 
disagreement in the ratings are irrelevant, while agreement becomes dominant as $\alpha\to\infty$.

To test our prediction scheme, we select a subset of the dataset comprised of $N$ users and 6000 movies (the parameters are listed in Table 1).  For varying values of $N$ and $\alpha$, we choose $k$ users from the data set in order to test the efficacy of our approach ($k$ is shown in the third 
column in Table 1) .  For each node $i$ selected, we iteratively perform the followings steps for each movie $l$ user $i$ has seen:  (I) remove the rating user $i$ gave to movie $l$ from the 
network, (II) compute the Erd\"os numbers for this modified network using (\ref{EnumCalc}), and (III) compute the predicted rating user $i$ gives to movie $l$ using (\ref{predictEq}) as a function 
of $\beta$, $p_i^{(l)}(\beta)$.  The average improvement as a function of $\beta$ is determined from the RMSD $\r_i^2(\beta)=\sum_l [r_i^{(l)}-p_i^{(l)}]^2/n_i$, where $n_i$ is the number of 
movies that user $i$ has seen.

The RMSD $\rho_i$ depends strongly on the number of movies  ($n_i$) that the user has seen, as can be seen by computing the average RMSD restricted to users with $n_i\ge n_{min}$.  In 
the upper inset of Fig.\ref{RMSD.fig}, a pronounced minimum in $\la \rho(\beta)\ra$ occurs for increasing $n_{min}$.  The relative improvement of  (\ref{predictEq}) over an unweighted average 
($\kappa_\beta=1-\la \r(\beta)\ra/\la \r(0)\ra$) is significant for $n_{min}\gtrsim 30$ as seen in the main panel of Fig. \ref{RMSD.fig}.  Restricting ourselves to users with $n_i\ge 30$ ratings gives 
an improvement of at least 3-5\% at $\beta=2$ for all values of $\alpha$ and $k$ examined (over 50\% of the nodes included in the average, see Table 1).  For very well connected nodes (with 
$n_{min}=200$ or about 8\% of the nodes in each case, see the lower inset of Fig. \ref{RMSD.fig}) the average improvement is quite significant, ranging from 4.5-9.5\%.  The dependence of the 
improvement on $n_{min}$ is somewhat unsurprising, as the preferences of users who have seen very few movies will be much more difficult to predict.  We also note that the negative 
improvement for small $n_{min}$ is due to the fact that the positions of the minimum in $\la\rho(\beta)\ra$ saturate at $\beta=2$ for large $n_{min}$, but are far from this value for small 
$n_{min}$.

Our minimal definition of the Generalized Erd\"os number which arises from an asymmetric measure of `closeness'  takes the global topology of the network into account. We have shown that
it can be used to characterize connectivity on simple analytically tractable networks as well as the basis for a ranking scheme for data sets from the Netflix prize, where it outperforms baseline 
schemes.  The weighted average in  (\ref{predictEq}) can be implemented in other prediction schemes, and a more complex form for the weighting between nodes (incorporating temporal information, for example) may give further improvements in predictions.  A natural next step of any measure of connectedness is to to use it in additional applications: problems associated with community detection in graphs, as well as the dynamics of 
diffusion, epidemics and the behavior of dynamic networks  with time-dependent edge weights beckon.

{\em{Acknowledgements:}}  The authors thank M. Venkadesan for numerous useful discussions and the suggestion of examining the NetFlix data.

\end{document}